\documentstyle[12pt]{article}
\begin{document}
\newcommand{\be}{\begin{equation}}
\newcommand{\ee}{\end{equation}}
\begin{titlepage}
\title{Singularities in scalar-tensor gravity}
\author{Valerio Faraoni    \\ 
           \\{\small \it  Physics Department, University 
of Northern British Columbia} \\ 
{\small \it 3333 University Way, Prince George, B.C., Canada V2N 
4Z9}}
\date{
} 
\maketitle
\thispagestyle{empty}
\vspace*{1truecm}
\begin{abstract}
The analysis of certain singularities in scalar-tensor gravity contained in a recent paper is 
completed, and situations are pointed out in which these singularities cannot occur.
\end{abstract}
\vspace*{0.5truecm}  
\end{titlepage}  \clearpage

Recently, Abramo {\em et al.} \cite{Abramoetal03} have investigated certain geometric 
singularities occurring in the gravitational sector of scalar-tensor theories described by the 
action
\be \label{action}
S=\int d^4 x \, \sqrt{-g} \left[ \frac{f(\phi)}{2} \, R -\frac{ \omega( \phi )}{2} \, 
g^{ac}\nabla_a \phi \, \nabla_c \phi -V( \phi) \right]  \;.
\ee
Their analysis generalizes to scalar-tensor gravity previous studies of nonminimally coupled 
scalar field theory. It is customary to interpret 
\be
G_{eff}( \phi )  \equiv \frac{1}{8\pi f( \phi)} 
\ee
as an effective gravitational coupling. However, in the weak field limit,  the 
effective coupling measured in a Cavendish experiment assumes the different form 
\cite{Nordtvedt68,footnote0}
\be \label{Gstar}
G^*_{eff}( \phi ) = \frac{ 2\omega f +\left( 2df/d\phi
\right)^2}{ 8\pi f \left[ 2\omega f +3 \left( df/d\phi \right)^2 \right] } \;.
\ee
Both  effective couplings diverge at the zeros of  $f(\phi) $ and change sign when $\phi$ 
crosses a value corresponding to a zero of $f$.  Moreover, the conformal transformation  to 
the Einstein frame 
\be \label{newmetric}
g_{ab} \longrightarrow \tilde{g}_{ab} = \sqrt{ f(\phi) } \, \, g_{ab} \;,
\ee

\be \label{newfield}
\tilde{\phi} = \int \frac{d\phi}{f(\phi)} \, \sqrt{ f(\phi) + \frac{3}{2} \left( 
\frac{df}{d\phi} \right)^2} \;,
\ee
degenerates when $f(\phi)=0$. If $f(\phi) \leq 0 $,  is it in principle possible  
that also
\be
f_1( \phi) \equiv f(\phi) +\, \frac{3}{2} \left( \frac{df}{d\phi}
\right)^2
\ee
vanish \cite{footnote1}, which causes the effective coupling (\ref{Gstar}) to diverge if 
$\omega=1$ (the case considered by the authors of Ref.~\cite{Abramoetal03}). 
The set of points at which  $f_1(\phi)=0$ constitutes a singularity in
the transformation (\ref{newfield}) that is necessary to bring 
the  scalar field into canonical form in  the Einstein frame.

The first kind of singularity is usually avoided  by 
imposing that $f(\phi)>0 $, which guarantees that gravity is 
attractive and that the graviton carries positive energy. However, the scalar may be  allowed 
to assume values that make $f(\phi) $
negative, which are not {\em a priori} forbidden by the classical
dynamics. This is the case, e.g.,  of the nonminimally coupled theory described
by  $f(\phi)=\left( 8\pi G \right)^{-1}-\xi \phi^2$ --  in the  literature one often 
encounters situations in 
which  $ G_{eff}\equiv G \left( 1-8\pi G \xi \phi^2 \right)^{-1} < 0 $ 
\cite{Gurevichetal73}. This regime is 
sometimes disguised by the fact that the field equations are written as
\be \label{fieldequation}
G_{ab}=8\pi \, G_{eff} \, T^{(eff)}_{ab}\left[ \phi \right] \;,
\ee
where $ T^{(eff)}_{ab}\left[ \phi \right] $ is an effective
energy-momentum tensor for the scalar $\phi$ \cite{footnote2}. When the time-time component of 
eq.~(\ref{fieldequation}) is considered, $ G_{eff}$ is allowed to become negative but
the Hamiltonian constraint is satisfied because the effective
energy density of the scalar $T^{(eff)}_{ab}\left[ \phi \right] \, u^a\,
u^b $ (where $u^a $ denotes the time direction) also
becomes negative.

The singularity $f(\phi)=0$ in $ G_{eff}$ corresponds to the loss of
predictability at an hypersurface that is not hidden inside an event horizon. The Cauchy 
problem is well-posed in the
Einstein frame, but not in the Jordan frame \cite{TeyssandierTourrenc83} and the vanishing of 
$f$ precludes the
possibility of defining the Einstein frame metric (\ref{newmetric}). Although even in 
general relativity not every 
solution corresponds to a globally
hyperbolic spacetime (e.g., $ pp$-waves \cite{Penrose65}), for many
authors  the idea of a naked Cauchy horizon is sufficiently unpleasant  
to require {\em a priori} $f( \phi) >0$. An added disadvantage of the $f=0$ singularity
is that it invalidates the covariant and gauge-invariant analysis of cosmological 
perturbations \cite{Hwang}   and the equations ruling the evolution of these perturbations 
become singular when $f\rightarrow 0$.  Nevertheless, the dynamics
of the unperturbed universe do not disallow values of $\phi $ such that $ f(\phi)=0$. 
Abramo {\em et al.} have studied the singularities $f( \phi)=0 $ and $f_1(\phi)=0$ in a 
Bianchi~I model: both of them  turn out to be
true spacetime singularities with divergent Kretschmann scalar $R_{abcd} 
R^{abcd}$ \cite{Abramoetal03,Abramoetal03b}.

Let us turn our attention now to the second kind of singularities 
$f_1(\phi)=0$. For this singularity to occur it must be $f(\phi) \leq 0 $  and hence 
requiring  $f(\phi) $ to always be positive eliminates 
this singularity also. Further, even if negative values of $f$ are 
permitted, the singularity $f_1( \phi)=0 $ may be forbidden by the classical dynamics
(this is the case, e.g., of closed or critically open Friedmann-Lemaitre-Robertson-Walker 
(hereafter ``FLRW'') universes under reasonable
assumptions --- see below).  In addition to the divergence of the effective coupling 
(\ref{Gstar}) if $\omega=1$, the $f_1=0$ singularity in
eq.~(\ref{newfield}) precludes the possibility of reformulating the scalar-tensor theory
in the Einstein frame. This circumstance does not {\em a priori} imply
any pathology in the dynamics, but the $f_1=0$ singularity leaves one
uncomfortable for the following reasons:
\begin{enumerate}

\item The Cauchy problem is not well-posed in the Jordan
frame \cite{TeyssandierTourrenc83}.

\item In the quantization of linearized scalar-tensor  gravity, it is the Einstein
frame and not the Jordan frame metric perturbation that is
identified with the physical graviton and carries positive energy 
\cite{Soleng88b}-\cite{DamourFarese92}. Being  
unable to  define the Einstein frame then means that one cannot  quantize 
linearized gravity.

\end{enumerate}

Abramo {\em et al.}  \cite{Abramoetal03,Abramoetal03b} have shown that 
$f_1(\phi)=0$ occurs  unless two conditions are met: $ f $ and $ df/d\phi $
simultaneously vanish and $ f_1$ has as zeros only the zeros of $f $ --- such a situation 
occurs if  $ f( \phi) \propto \phi^{2n} $ and $V( \phi ) \propto \phi^{2(2n-1)} $. This 
situation  is discarded as fine-tuned by Abramo {\em et al.} but it is not at all   
unphysical:  this is the case of induced gravity with a massive 
scalar, described by  $f(\phi)=\epsilon \phi^2$ and $ V(\phi) =m^2\phi^2 /2$.

Finally, we show that the $f_1=0 $ singularity is dynamically avoided in a closed or 
critically open FLRW  universe under physically reasonable assumptions. Our argument 
generalizes to scalar-tensor gravity an argument previously given in 
Refs.~\cite{FutamaseMaeda89,Amendolaetal90} for the special case of non-minimally coupled 
scalar 
field theory.

The  Hamiltonian constraint in a FLRW  universe, 
\be
3f \left( H^2+\frac{K}{a^2} \right)= \rho^{(m)} +\frac{\omega}{2} \left(
\dot{\phi} \right)^2 +V -3H \dot{f} \;,
\ee
where $\rho^{(m)} $ is the energy density of matter, can be rewritten as
\be  \label{birbabirbone}
\left( H + \frac{ \dot{f} }{ 2f} \right)^2 = \left( \frac{\dot{f}}{2f}
\right)^2 
+ \, \frac{ \omega \left( \dot{\phi} \right)^2}{6\,f} +
\frac{\rho^{(m)} }{3f} - \frac{K}{a^2 } +
\frac{V}{ 3f}   \;. 
\ee
Now let us assume that
$ f+\frac{3}{2} \left( \frac{df}{d\phi} \right)^2 \leq 0 $ (which implies
that $ f \leq 0 $) --- then the first two terms on the right hand side add up to 
\begin{eqnarray}
&& \left( \frac{\dot{f}}{2f} \right)^2 +\, 
\frac{ \omega \left( \dot{\phi} \right)^2}{6\, f} = 
\frac{1}{6} \, \left( \frac{ \dot{\phi}
}{f} \right)^2 \left[ \omega f +\frac{3}{2}\left( \frac{df}{d\phi}
\right)^2 \right]  \nonumber \\
&& <
 \frac{1}{6} \, \left( \frac{ \dot{\phi}
}{f} \right)^2 \left( \omega -1 \right) f \leq  0 \;,
\end{eqnarray}
where the last inequality is satisfied if $\omega \geq 1$, as is usually
assumed, and hence the first two terms on the right hand side of
eq.~(\ref{birbabirbone})  give non-positive contributions. Now, by
assuming that $V \geq
0$, $K=0$ or $ +1$, and $\rho^{(m)} \geq 0$, one also obtains 
\be
\left(  \rho^{(m)}   + V \right) \frac{1}{ 3f} - \frac{K}{a^2} \leq 0
\;,
\ee
with the equality holding only in Minkowski space. Hence the left hand side of
eq.~(\ref{birbabirbone}) is non-negative, while the right hand side is
negative, an absurdity caused by assuming 
that $ f_1 \equiv f+\frac{3}{2} \left( \frac{df}{d\phi} \right)^2 \leq 0 $. 
The assumptions used in our proof automatically guarantee that $f_1(\phi)> 0 $ and that the 
$f_1=0$ singularity is dynamically avoided.

The consideration of a FLRW universe is not very relevant before inflation because the initial 
conditions are likely to be anisotropic, and it is inflation that leads  to a highly  
homogeneous and isotropic universe. However, a FLRW metric is mandatory in  quintessence  
models of the universe during the present era of accelerated expansion, 
where the conditions of this comment apply. 

Although the interest of Ref.~\cite{Abramoetal03} in singularities of scalar-tensor 
gravity arises from nonminimally coupled scalar field cosmology, $f=0$ singularities  
also appear in studies of wormholes with nonminimally coupled scalar fields 
(\cite{BarceloVisser00}  and references therein). In the future it would be interesting to 
study the stability of wormhole  solutions corresponding to $f=0$.



\begin{thebibliography}{99}
\bibitem{footnote0} This expression can also be derived from cosmological perturbation theory
\cite{Boisseauetal00}.
\bibitem{footnote1} In $D>2$ spacetime dimensions the function $f_1$ is defined as
$$
f_1( \phi) \equiv f(\phi) +\, \frac{D-1}{D-2} \left( \frac{df}{d\phi}
\right)^2 \;.
$$
\bibitem{footnote2} There are three possible and physically inequivalent ways of writing the 
field equations for nonminimally coupled scalar fields, and care must be taken in choosing the 
appropriate one \cite{BellucciFaraoni02}.
\end{thebibliography}
\end{document}